\documentstyle[preprint,aps]{revtex}
\begin{document}
\draft
\preprint{UCF-CM-93-007}
\title
{Density matrix for a mesoscopic current ribbon}
\author{O. Heinonen and M.D. Johnson}
\address{
Department of Physics, University of Central Florida, Orlando, FL 32816-2385
}
\maketitle
\begin{abstract}
We consider an ideal mesoscopic ribbon in which a steady azimuthal
current is generated.
We show that the closed interacting electron
system in the presence of the current is described by a density matrix
which is that
of an {\em equilibrium} system {\em without} current but with a constrained
Hamiltonian.
\end{abstract}
\pagebreak

\narrowtext
Two recent papers present new approaches to nonequilibrium steady-state
systems. The important common point made in both papers is that
steady-state mesoscopic systems can be described by a density matrix
which has the form of an {\em equilibrium} density matrix, but with
a constrained Hamiltonian. In
the first paper, Hershfield \cite{Hershfield} demonstrated that the
density matrix $\hat\rho$ of a steady-state nonequilibrium quantum
system has the general form
\begin{equation}
\hat\rho=e^{-\beta(\hat H-\hat Y)},\label{rho_Selman}
\end{equation}
where $\hat H$ is the Hamiltonian, $\beta$ is the inverse temperature
(we will use units in which $k_B=\hbar=e=m_e=1$),
and $\hat Y$ is an operator which depends on
how the system is driven out of equilibrium. This operator is
defined implicitly in terms of a non-trivial infinite
set of differential equations.
In the other paper (Ref.~\onlinecite{HJ1}, hereafter referred to as HJ),
we formulated an approach to
steady-state mesoscopic transport based on the maximum entropy principle
of nonequilibrium statistical mechanics \cite{Jaynes}. Using the
maximum entropy principle, we derived the following density matrix
for a multi-terminal steady-state mesoscopic system:
\begin{equation}
\hat\rho=e^{-\beta\left(\hat H-\mu \hat N-\sum_\alpha
\xi_\alpha \hat I_\alpha\right)},
\label{rho_HJ}
\end{equation}
where $\hat I_\alpha$ is the current incoming from terminal $\alpha$ and
the Lagrangian multipliers $\xi_\alpha$ are adjusted to give
the applied source-drain current and zero current at all other terminals.
This density matrix, like that in Eq.~(\ref{rho_Selman}),
has the form of an {\em equilibrium} density matrix
of a constrained Hamiltonian. It can also be argued on general grounds
using Galilean invariance that distributions of the form obtained
from Eq.~(\ref{rho_HJ}) can be expected
in an infinitely long ideal mesoscopic wire \cite{HJ1,Hajdu}.

One important point to be noted is that the single-particle distributions
obtained in HJ \cite{HJ1,distributions} differ, even when linearized with
respect to the currents, from the local-equilibrium distributions
typically used in
the Landauer-B\"uttiker \cite{LB} formalism of
mesoscopic transport. In the linear-response regime, one would not
expect that measurements of, for example, conductivity directly
probe the electron distributions, since such quantities can in the
linear-response approximation be
expressed as the trace over the {\em equilibrium} density
matrix and two-point correlation functions. However, experiments
conducted beyond the linear-response regime should be sensitive to
the nonequilibrium steady-state electron distributions. This should be the
case for precision measurements of the integer quantum Hall effect,
where the measured Hall voltage exceeds $16\hbar\omega_c/e$, with
$\omega_c$ the cyclotron frequency.  We demonstrated in HJ
that a transport theory based on Eq.~(\ref{rho_HJ}) can explain
the quantization of the integer quantum Hall effect at such large, but
experimentally typical, currents, while the Landauer-B\"uttiker formalism
fails to do so.

In deriving the density matrix Eq.~(\ref{rho_HJ}) using the maximum entropy
principle, it was necessary to make assumptions about which are the
relevant observables, the expectation values of which are taken to be known.
While this is a standard procedure of the maximum entropy approach to
statistical mechanics,
it is also the source of some controversy about this approach,
since no unambiguous procedure for choosing the
observables exists \cite{Landauer}.
It is therefore desirable to inquire whether such steady-state distributions
in mesoscopic systems can be obtained by other means.
The purpose of the present paper is to present one such
example. We will show by an explicit
calculation that the exact density matrix
in the presence of a steady current is precisely that of HJ
for a simple specific case: a closed system of interacting
(spinless) electrons in
an ideal, mesoscopic, two-dimensional ribbon.
Here we obtain this result by considerations of equilibrium thermodynamics
in a rotating reference frame, plus adiabatic switching-on, without appeal
to the maximum entropy principle.
By an ideal system we mean that there are no elastic or inelastic
scattering processes other than those resulting from electron-electron
interactions. A current is generated in the electron system by
adiabatically threading the ribbon with an integral number of flux quanta.
In this
case, there is then a one-to-one correspondence
between the many-body eigenstates in the presence and in the absence of
the current. This allows us to demonstrate that the density matrix
in the presence of the current
can be directly related to the density matrix in the absence of
current, and has the form of an {\em equilibrium}
density matrix with a constrained Hamiltonian.
While we draw no conclusions about
{\em open} systems here, we provide an explicit
example of a system which by an independent calculation proves to be
described by the density matrix obtained by HJ.

The system considered here resembles those considered
in investigations of persistent currents \cite{persistent}
in mesoscopic rings.
The important difference is that in those investigations the
current as a function of an applied AC magnetic flux is
typically calculated for
a non-interacting system in the presence of elastic scatterers. (The
role of interactions in the presence of disorder is complicated and unclear.
\cite{unclear})
Great care has to be taken to ensemble-average correctly, and
to correctly account for the magnetic field actually penetrating the
ring itself in an experiment. Also, in the presence of scatterers which
break the rotational invariance there is no adiabatic
curve-crossing as the magnetic flux is increased adiabatically.
Here we are considering a closed, impurity-free
{\em interacting}
electron system, and the flux is used only as a device \cite{Kohn} to generate
the electric field and the resulting current. Our ultimate
goal is to obtain the exact density matrix in the presence of a steady current.

We take the system to be a two-dimensional interacting electron
gas confined to a cylindrical ribbon of radius $R$.  Positions on the ribbon
are described by cylindrical coordinates $(r,\theta,z)$, with $r=R$.  A uniform
azimuthal electric field $E\hat\theta$ is generated by
piercing the bore of the cylinder with adiabatically increasing magnetic flux.
The electric field
is described by a time-dependent vector potential
\begin{equation}
{\bf A}(r,\theta;t)={\Phi\over 2\pi r}f(t)\hat\theta,\label{vector_A}
\end{equation}
where $\Phi f$ is the magnetic flux piercing the cylinder's bore.
The monotonically non-decreasing function
$f(t)$ describes the adiabatic turning-on of the vector
potential, with $f(t\to-\infty)=0$ and $f(t\to\infty)=1$.
We will assume that $\Phi=p\Phi_0$
with $p$ an integer and $\Phi_0=2\pi c$ the flux quantum,
so that the cylinder contains an integral number of
flux quanta as $t\to\infty$. From
\begin{equation}
{\bf E}=-{1\over c}{\partial {\bf A}\over\partial t}
\label{el}
\end{equation}
the electric field then vanishes for $t\to\pm\infty$ but is finite during
the time that $f$ is changing.  This finite electric field sets up
an azimuthal current which persists as the electric field vanishes,
since the system is dissipationless.

With the vector potential given by Eq. (\ref{vector_A}),
the first-quantized many-body Hamiltonian is
\begin{equation}
H=\sum_i{1\over 2m^*}\left[ -{\partial^2\over \partial z_i^2}
+{1\over R^2}\left(L_{z,i}+pf(t)\right)^2\right]
+\sum_iV_c(z_i)+\frac{1}{2}\sum_{i\not=j}V(r_{ij}),\label{H1}
\end{equation}
where the sums are over the $N$ particles of the system, $V_c(z)$ is a
confining potential, $r_{ij}=|{\bf r}_i-{\bf r}_j|$, and
$V(r)$ is the electron-electron interaction.
Note that $L_{z,i}=-i\partial/\partial\theta_i$ is the canonical
angular momentum operator, which is independent of $\bf A$.
The eigenstates of $H$ are given by $N$-particle wavefunctions
$\Psi=\Psi(z_1\theta_1, z_2\theta_2,\ldots,z_N\theta_N)$.
To proceed, we second-quantize the Hamiltonian Eq. (\ref{H1}) using the
field operators
\begin{equation}
\hat\psi(z,\theta)=\sum_{mn}c_{mn}\psi_{mn}(z,\theta)
\label{fields}
\end{equation}
where the single-particle wavefunctions
$\psi_{mn}(z,\theta)={1\over\sqrt{2\pi R}}e^{im\theta}\phi_{mn}(z)$
(normalized to give probability per unit area) satisfy
\begin{equation}
\left[-{1\over 2m^*}\nabla^2 +V_c(z)\right] \psi_{mn}(z,\theta)
=\epsilon_{mn}\psi_{mn}(z,\theta).\label{single_particle}
\end{equation}
Here $n$ is a subband index and $m$ an angular momentum index.
The second-quantized Hamiltonian is
\begin{eqnarray}
\hat H&=&\sum_{mn}\epsilon_{mn}c^\dagger_{mn}c_{mn}^{\phantom{\dagger}}
+\sum_{mn}{\left[pf(t)\right]^2+2mpf(t) \over 2 m^*R^2}c^\dagger_{mn}
c_{mn}^{\phantom{\dagger}}
\nonumber \\
&&+\frac{1}{2}\sum_{{mm'k}\atop{n_1 n_2 n_3 n_4}}
V_{k;n_1,n_2,n_3,n_4}c^\dagger_{m,n_1}c^\dagger_{m'-k,n_2}
c_{m',n_3}^{\phantom{\dagger}}c_{m-k,n_4}^{\phantom{\dagger}},
\label{H2}
\end{eqnarray}
with $V_{k;n_1,n_2,n_3,n_4}$ the matrix elements of
the particle interaction.
The total angular momentum operator $L_z=\sum_iL_{z,i}$ becomes
\begin{equation}
\hat L_z=\sum_{m,n}m c^\dagger_{mn}c_{mn}^{\phantom{\dagger}}.\label{L_theta}
\end{equation}
The azimuthal current density operator is
$\hat j_\theta(z,\theta)={1\over2}[\hat\psi^{\dagger} v_{\theta}\hat\psi +
(v_{\theta}\hat\psi)^{\dagger}\hat\psi ]$, where
$v_{\theta}={1\over2m^{*}R}(L_z+pf)$ is the speed in the azimuthal direction.
This can be written
\begin{equation}
\hat j_\theta(z,\theta)={1\over2\pi m^*R^2}\sum_{mm',nn'}
\left[{m'+m\over 2}+pf(t)\right]
e^{-i(m-m')\theta}\phi^*_{mn}(z)\phi_{m'n'}^{\phantom{*}}(z)
c_{mn}^\dagger c_{m'n'}^{\phantom{\dagger}}\label{j_theta}.
\end{equation}
We also introduce an azimuthal current operator
$\hat I$ by integrating $\hat j_{\theta}(z,\theta)$ over $z$
and averaging over $\theta$, which yields
\begin{equation}
\hat I={1\over 2\pi m^* R^2}\sum_{mn}[m+pf(t)]
c_{mn}^\dagger c_{mn}^{\phantom{\dagger}}.
\label{I_op}\end{equation}
It is straightforward to show that
$\hat H$, $\hat I$, $\hat L_z$, and the number operator
$\hat N=\sum_{mn}c_{mn}^{\dagger}c_{mn}^{\phantom{\dagger}}$ all commute,
so we choose a basis in which all are diagonal.  In fact,
\begin{equation}
\hat I = {1\over2\pi m^{*}R^2}(\hat L_z + pf \hat N),
\label{I_op2}
\end{equation}
so a basis diagonal in $\hat L_z$ and $\hat N$ is automatically diagonal in
$\hat I$, and the eigenvalues $I$ and $M$ of $\hat I$ and $\hat L_z$ are
simply related.

Let us consider next the adiabatic evolution of some $N$-electron
wavefunction $\Psi$.  Suppose that $\Psi$ is an eigenstate of
$H_0\equiv H(f=0)$ with energy $E$ and total angular momentum $M$. From
Eq.~(\ref{H1}), the Hamiltonian in the presence of the flux tube can
be written
\begin{equation}
H= H_0 + {pf\over m^{*}R^2}\left(L_z + {1\over2}pf N\right).
\label{H3}
\end{equation}
As long as the field is switched on adiabatically ({\it i.e.},
$df/dt$ is much smaller than electron-electron scattering rates),
$\Psi$ remains an eigenstate of $H$ when $f\neq0$.
Although its (canonical) total angular momentum is unchanged,
its energy and current change with $f$.  Since $p$ is an integer,
the vector potential as $f\rightarrow1$ can be removed by a unitary
gauge transformation:
\begin{eqnarray}
\widetilde \Psi &=& e^{ip\sum_i\theta_i}\Psi,\nonumber\\
\widetilde H &=& e^{ip\sum_i\theta_i}H e^{-ip\sum_i\theta_i}=H_0 .
\label{unitary}
\end{eqnarray}
The transformed state $\widetilde \Psi$ has total angular momentum
$M+pN$, and is an eigenstate of $H_0$.  Thus, the spectrum of
the Hamiltonian for $t\to\infty$ is identical to the spectrum for
$t\to-\infty$.  More, the energy and current when $f=1$ of a state with
angular momentum $M$ are precisely those of an eigenstate
of $H_0$ with angular momentum $M+pN$.  It is clear from this and
Eq.~(\ref{H3}) that as $f$ grows from 0 to 1 the subset
of the spectrum consisting of eigenvalues of
states with angular momentum $M$ evolves
to the subset of eigenvalues of eigenstates of $H_0$ with angular
momentum $M+pN$.
This occurs with
no level crossings between states in this subspace (although crossings
occur between states with different $M$).

In particular, suppose we start at $t\to-\infty$
with the system in the interacting ground state (with
$M=0$) and adiabatically turn on the vector potential.
Then at $t\to\infty$ the system's energy and current will be that of
the lowest-energy eigenstate of $H_0$ with angular momentum $p N$.
Thus, the final energy and current are obtained by finding
the lowest-energy eigenstate of
$H_0$ in the subspace with $\langle \hat L_z\rangle=p N$. In
other words, the final state when a current has been turned on can
be found by extremizing $H_0$ subject to the constraint that
$\langle \hat L_z \rangle =pN$.
Using Eq.~(\ref{I_op2}), we can instead constrain the current to
be $\langle \hat I_0 \rangle = pN/2\pi m^{*}R^2$ (where $\hat I_0$ is the
current operator with $f=0$).
In practice, a convenient way to satisfy the constraint is to
introduce a Langrangian multiplier, and to look for stationary states of
\begin{equation}
\hat H_0-\xi\hat I_0\label{constraint1}
\end{equation}
where $\xi$ is chosen so that $\hat I_0$ has the required eigenvalue.

We next turn to finite temperatures, and consider a
system initially
in contact with a particle reservoir, which maintains
the average number of particles at $N$, and in thermal contact with a heat
reservoir at temperature $T$. Initially ($t\to-\infty$) the density matrix
is
\begin{equation}
\hat \rho_0 = e^{-\beta\left(\hat H_0 - \mu_0 \hat N\right)},
\label{rho0}
\end{equation}
and the system has zero average current,
$\langle\hat I_0\rangle_T\equiv Tr(\hat I_0 \hat\rho_0) /Tr(\hat\rho_0) =0$.
We then insulate the system thermally from the heat reservoir and adiabatically
turn on the vector potential while keeping the
average number $N$ fixed.  There are two ways to understand what
then happens.  First, notice that when the system is isolated thermally
each subspace of states with a given total angular momentum $M$ becomes
a separate subsystem.  Although electron-electron interactions could
cause re-equilibration within a subset, these interactions conserve
total angular momentum and so cannot, {\it e.g.}, transfer energy from
one angular momentum subspace to another.
Moreover, as explained above, the energy
of each state in a given subset changes by exactly the same amount when $f$
varies (see Eq.~(\ref{H3})).  That is, the spacings in energy between the
states in each subspace remain constant, and consequently the
occupancies of these states will not change.  Then the occupancy of a
state at $t\to\infty$ is given precisely by its initial occupancy at
$t\to-\infty$.  Consider an $N$-electron state with initial energy $E$
and angular momentum $M$.  Initially the occupancy of this
state is $e^{-\beta(E-\mu_0 N)}$.  At $t\to\infty$, this energy and
current of this state become
\begin{eqnarray}
E' &=& E + {p \over m^{*} R^2} (M+{1\over2}pN), \nonumber\\
I' &=& {1\over2\pi m^{*} R^2}(M+pN).
\label{evolved}
\end{eqnarray}
The occupancy of this state is still $e^{-\beta(E-\mu_0 N)}$.  That is,
the occupancy is determined not by its energy $E'$, but
by $E$, which can be related to $E'$ and $I'$ using the above.
Since the initial and final sets of states are identical (for $p$ an
integer), the density matrix can be written in terms of the $f=0$
operators $H_0$ and $I_0$.  As $t\to\infty$, the density matrix evolves
to
\begin{eqnarray}
\hat\rho &=& \exp\left\{-\beta\left(\hat H_0 - {p\over m^{*}R^2}\left(\hat L_z
+
{1\over2} p\hat N\right) - \mu_0 \hat N\right)\right\}\nonumber\\
&=& \exp\left\{-\beta\left(\hat H_0 - \xi \hat I_0 - \mu \hat N
\right)\right\},
\label{rho_evolve}
\end{eqnarray}
where the final equality defines $\xi=2\pi p$ and $\mu=\mu_0 + p^2/2m^{*}R^2$.

A second way to understand this result is to consider equilibrium in a rotating
reference frame.  As the vector potential
is turned on, the system evolves adiabatically as $t\to\infty$ to a state with
average current $\langle \hat I\rangle_T =I$.  After the electric field
has returned to zero (as $t\to\infty$), we transform
to a coordinate frame rotating with an angular velocity
$\Omega$ relative to the
lab frame, where $\Omega$ is chosen so that the azimuthal
current is zero in the rotating frame. The Hamiltonian ${\tilde H}$
in the rotating frame and the Hamiltonian $\hat H$ in the original frame
are related by \cite{LandL}:
\begin{equation}
{\tilde H}=\hat H -\Omega \hat L_z,
\label{omega_energy}
\end{equation}
where
\begin{equation}
\Omega={p\over m^*R^2}.\label{Omega}
\end{equation}

In the rotating frame, the system consists of an isolated
interacting electron gas at zero net current.
In this frame, one should expect
the electron-electron interactions to equilibrate the
system.  (This equilibration is at the same temperature $T$ that
the system was in at $t\to-\infty$ in the lab frame, since
the center-of-mass motion of a system does not change the temperature
\cite{Huang}.)
Thus, in the rotating frame, the system has an
{\em equilibrium} density matrix
\begin{equation}
{\tilde\rho}=e^{-\beta\left({\tilde H}-
\tilde \mu\hat N\right)}.
\label{tilde_rho}
\end{equation}
We express this density operator in terms of the Hamiltonian $\hat H_0$
of the stationary frame at zero flux by first using Eq.~(\ref{omega_energy})
and then gauging away the vector potential as above. With the operators in
their first-quantized form, this gauge transformation gives
\begin{equation}
e^{-\beta\left(H-\Omega\hat L-\tilde\mu N\right)}
\rightarrow e^{ip\sum_i\theta_i}e^{-\beta\left(
H-\Omega L-\tilde\mu N\right)} e^{-ip\sum_i\theta_i}
=e^{-\beta\left( H_0-\mu N-\Omega L\right)},\label{unitary2}
\end{equation}
where $\mu={\tilde\mu}+p^2/m^{*}R^2$. Thus the system is
described by the density matrix
\begin{equation}
\hat\rho=e^{-\beta\left(\hat H_0-\xi\hat I_0-
\mu\hat N\right)},\end{equation}
where we have used Eq.~(\ref{I_op2}) (with $\hat I_0=I(f=0)$), and as before
$\xi=2\pi p$.
In practice one can regard $\mu$ and $\xi$ as parameters which
are adjusted to give the required thermal averages of particle number and
current. Thus the density matrix is that of an equilibrium systems
described by a {\em constrained} Hamiltonian, and is identical to that
(Eq.~(\ref{rho_HJ}))
obtained in HJ by maximizing the entropy subject to constraints on
internal energy, particle number and total current.

This work was supported by the National Science Foundation through
grant DMR93-01433.

\end{document}